\documentclass[useAMS,usenatbib]{mn2e}

\usepackage{aas_macros}
\usepackage{multirow}
\usepackage{placeins}
\usepackage{url}
\usepackage{amssymb}
\usepackage{rotating}

\newcommand{\Msun}{\ensuremath{\textrm{M}_{\sun}}}

\title[Westerlund~1: a collapsing cluster with no mass segregation?]{Hierarchical formation of Westerlund~1: a collapsing cluster with no primordial mass segregation?\thanks{Based on observations collected at the European Southern Observatory, La Silla, Chile, and retrieved from the ESO archive (Program ID 67.C-0514)}}
\author[M. Gennaro et al.]{Mario Gennaro$^{1,2}$\thanks{E-mail:
gennaro@stsci.edu (MG)}, Simon P. Goodwin$^3$, Richard J. Parker$^3$\thanks{Royal Society Dorothy Hodgkin Fellow}, Richard J. Allison$^{4}$, \newauthor Wolfgang Brandner$^{2}$\\
$^{1}$Space Telescope Science Institute, 3700 San Martin Drive, Baltimore, MD 21218, USA\\
$^{2}$Max-Planck-Institut f{\"u}r Astronomie, K\"{o}nigstuhl 17, 69117, Heidelberg, Germany\\
$^{3}$Department of Physics and Astronomy, University of Sheffield, Sheffield, S3 7RH, UK\\
$^{4}$Zentrum f\"ur Astronomie der Universit\"at Heidelberg, Institut f\"ur Theoretische Astrophysik, \\ \hspace{1ex}Albert-Ueberle-Str.~2, 69120 Heidelberg, Germany\\
}

\begin{document}

\date{Accepted for publication in the MNRAS main journal on August 11 2017}

\pagerange{\pageref{firstpage}--\pageref{lastpage}} \pubyear{2011}

\maketitle

\begin{abstract}
We examine the level of substructure and mass segregation in the massive, young cluster Westerlund~1.  We find that it is relatively smooth, with little or no mass segregation, but with the massive stars in regions of significantly higher than average surface density. While an expanding or bouncing-back scenario for the evolution of Westerlund 1 cannot be ruled out, we argue that the most natural model to explain these observations is one in which Westerlund~1 formed with no primordial mass segregation and at a similar or larger size than we now observe.
\end{abstract}


\maketitle
%
%________________________________________________________________

\label{firstpage}

\section{Introduction}

How do stars form?  The answer to this apparently simple, but crucial,
question is still unclear.  We know stars form in dense molecular clouds, in numbers from a few up to millions of objects.  
Observations find that after about 10 Myr, roughly 10 per cent of
these stars are in  
bound (potentially long-lived) star clusters \citep{2003ARA&A..41...57L,2010RSPTA.368..713L}.   However, it is unclear if stars predominantly form in dense
clusters which are then destroyed, whether they form in low-density hierarchies in which some collapse to form clusters, or a mixture of both of these mechanisms. 

In the `clustered' star formation scenario, most stars form in dense, bound clusters, but 
most of these clusters are destroyed by gas expulsion \citep[e.g.][]{1978A&A....70...57T, 1984ApJ...285..141L,2006MNRAS.373..752G, 2007MNRAS.380.1589B}, or by their birth 
environments \citep[e.g.][]{2011MNRAS.414.1339K}.  In the scenario for `hierarchical' star formation
stars form in complex hierarchical distributions and a small fraction
are locally bound and able to collapse 
to form star clusters \citep[e.g.][]{2002Ap&SS.281...51A,2006A&A...445..979P,2009ApJ...700L..99A,2010MNRAS.409L..54B,2014MNRAS.445..674C,2014prpl.conf..291L}.

The differences between these two possible extremes of star formation are important. 
In clustered star formation almost all stars spend time in dense
environments which will alter their multiplicity, and can perturb
discs and affect planet formation \citep{1995MNRAS.277.1491K,2004ApJ...611..360A,2012MNRAS.419.2448P,2015A&A...577A.115V}.  However, in hierarchical star formation a significant
fraction of stellar systems could avoid interactions and disperse
almost unaltered into the field.

The density at which stars form could also influence the way in which massive stars form \citep[e.g.][]{2007ARA&A..45..481Z}.  In clustered star formation massive 
stars would almost certainly form
via competitive accretion as there is not enough space between stars to form
individual massive cores \citep[e.g.][]{1997MNRAS.285..201B}.  On the contrary, in
hierarchical star formation massive stars would be expected to form
monolithically \citep[e.g.][]{2007ApJ...656..959K} as there is no massive
cluster in place at their formation in which they can assemble.  

The feedback from massive stars into the galactic environment is also
enhanced in dense star clusters \citep[where the most massive stars can
`gang-up',][]{2011ApJ...731...91L}.  Therefore the impact of massive stars
on larger scales could change significantly depending on their typical
birth environment. 

Detailed observations of young star clusters should help distinguish
between formation models.  In both clustered and hierarchical star
formation there should be around 10 per cent of stars in $\sim 1$~pc
clusters after around 5--10~Myr in order to fit the observations.  In
clustered star formation we expect that the cluster has always been
dense, and probably denser in the past as it has at some point had to
expel its residual gas \citep[][]{2015MNRAS.447..728B}.  In hierarchical star
formation the cluster may have been less dense in the past and has collapsed
down to its current state \citep[e.g.][]{2009ApJ...700L..99A}.   

It is possible to place constraints the past state of a cluster/association by looking
at the distributions of the most massive stars (mass segregation and relative 
local densities) in tandem with the degree of structure in the cluster \citep{2014MNRAS.438..620P,2014MNRAS.438..639W, 2014MNRAS.445.4037P}. 

In this paper we examine the distributions of massive stars and
the structure of the massive cluster Westerlund~1 \citep[Wd~1,][]{1961AJ.....66T..57W}.  Wd~1 is a $\sim
5$~Myr old, $\sim 2$~pc across \citep[half-mass radius of $\sim 1$~pc, see][]{Brandner:2008lr}, and $\sim 5 \times 10^4$~\Msun~
cluster making it the most massive young cluster in the Galaxy \citep{2005A&A...434..949C}.   

The paper is organised as follows. In Sections~\ref{sec:data}, \ref{sec:catsel}, \ref{sec:models} and \ref{sec:indmass} we describe the data, catalogue selection, stellar models and determination of individual stellar masses. In Section~\ref{sec:structure} we present a structural analysis of Westerlund~1. In Section~\ref{sec:discuss} we provide a discussion, and we conclude in Section~\ref{sec:conc}.

\section{The Data}
\label{sec:data}

\begin{figure}
 \centering
	\resizebox{\hsize}{!}{\includegraphics{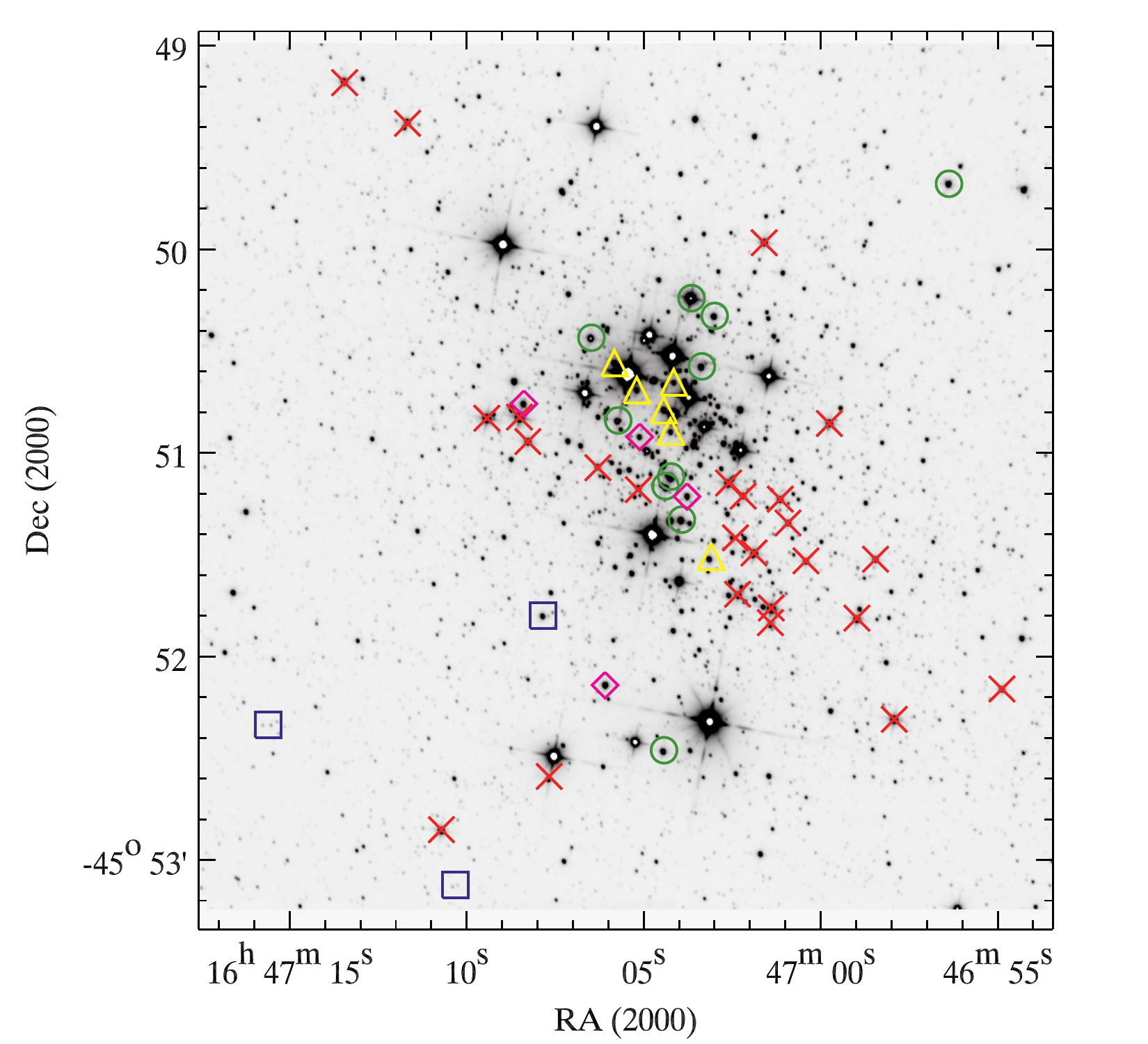}}
	\caption{SOFI $K_{\rmn{S}}$ band image of Wd~1. Superimposed
          are the position of 2MASS sources with $K_{\rmn{S}} \leq
          9.1$ (red crosses), the other isolated 2MASS sources with
          $K_{\rmn{S}} > 9.1$ mag (blue squares) and sources for which
          magnitudes were derived by PSF reconstruction (green
          circles). These are only the sources that survive the $3-\sigma$ clipping procedure in the near-infrared color-magnitude diagram described in Sect.~\ref{sec:catsel}. The position of the stars without near-infrared
          magnitude measurements but with optical magnitudes are
          displayed as magenta diamonds \citep[stars
            from][]{2007AJ....133.2696B} and yellow triangles
          \citep[stars from][]{Negueruela:2010fj}.} 
 \label{fig:spdistr}
\end{figure}

The data set used in the present analysis has been extensively
described in \cite{Brandner:2008lr} --hereafter B08-- and used also in
\cite{gennaro11} --hereafter G11. We refer the reader to these two
papers for a full description of the data reduction and analysis
process. In the following, we briefly summarize the reduction steps
and mainly emphasize the differences in the analysis that have been
introduced in the present work with respect to the previous two
papers. 

ESO NTT/SofI $J \, ,H \, \rmn{and} \, K_{\rmn{S}}$ broad band observations of Wd~1 (RA$_{\rmn{J2000}} = 16^{\rmn{h}}47^{\rmn{m}}03^{\rmn{s}}$, Dec$_{\rmn{J2000}} = -45\degr 50\arcmin 37\arcsec$) and of a nearby comparison field (offset by $\approx 7\arcmin$ to the East and $\approx 13\arcmin$ to the South of Wd~1), each covering an area of $4\farcm5 \times 4\farcm5$, were retrieved from the ESO archive (PI: J.Alves).
The reduction process is the same as in B08, performed using the
eclipse jitter routines \citep{Devillard:2001fk}. Point Spread
Function (PSF) fitting photometry was derived using the \texttt{IRAF}
implementation of \texttt{DAOPHOT} \citep{Stetson:1987qy}. 

The catalog of stellar objects was obtained by positional matching of the lists of $J \, ,H \, \rmn{and} \, K_{\rmn{S}}$ detections. The total number of stars is 6201 for the Wd~1 field and 4874 for the comparison field.
As in B08 and G11, we did not use the $H$ band measurements in the
analysis. The reason is that the largest colour baseline is attained
in the $J - K_{\rmn{S}}$ colour. Using this colour, the cluster
sequence in the colour-magnitude diagram (CMD) appears to be better
separated from the fore- and background contamination, thus providing
the most useful information for photometric member selection. Also the
larger colour baseline provides more reliable estimates for the
reddening value, giving better constraints on the average extinction
towards the cluster and therefore on the stellar masses, which we
infer from comparison of photometry and isochrones (see
Sect. \ref{sec:indmass}). 

There are some small differences in the photometric calibration
process compared to B08 and G11 though. Photometric zero points and
colour terms were computed by comparison of instrumental magnitudes of
relatively isolated, bright sources with counterparts in the 2MASS
Point Source Catalogue \citep[2MASS-PSC,][]{Skrutskie:2006uq},
similarly to B08. However we have refined the selection process for
the 2MASS reference stars and improved the fitting procedure for the
magnitude comparison.  
The results are very similar to B08, with an average difference in
zero-points of a few hundredths of magnitudes (the random uncertainty on the zero point being of the order of 1.5 thousandth magnitudes).  
The change in the calibration's colour term is similarly very small and, 
when limiting the comparison to the main sequence (MS) stars of Wd~1,
the $J-K$ colour with the current calibration is redder by
only 0.005 mag compared to B08 and G11.

We make our full photometric catalog available; the catalog contains
all the SofI detections with magnitudes calibrated in the 2MASS
system. 
The stars that survive the selection process (see
Sect.~\ref{sec:catsel}) and that are actually used in this work are
marked by additional flags. 
We also provide the photometric mass determinations and additional
information extracted from the 2MASS-PSC  \citep{2007AJ....133.2696B,Negueruela:2010fj}. An example of the information available
in the catalog is provided in Table~\ref{tab:cat}, the full version is
available online.

\begin{table*}
\begin{minipage}{175mm}
\centering
\renewcommand\tabcolsep{3.pt}
\begin{tabular}{ccccccccccccc}
\hline
\hline
RA & Dec & Source  & M$_{\rmn{NIR}}$ & M$_{\rmn{OPT}}$ &
$J_{\rmn{SofI}}$ & $K_{\rmn{SofI}}$ & $J_{\rmn{2M}}$ & $K_{\rmn{2M}}$ &
ID$_{\rmn{N}}^1$ \footnotetext{$^1$ From \cite{Negueruela:2010fj}} & Sp.Type$_{\rmn{N}}^1$ & ID$_{\rmn{B}}^2$ \footnotetext{$^2$ From \cite{2007AJ....133.2696B}} & Sp.Type$_{\rmn{B}}^2$ \\
  $[$deg$]$ & $[$deg$]$ & -- &  $[$\Msun$]$ & $[$\Msun$]$ &
$[$mag$]$ & $[$mag$]$ & $[$mag$]$ & $[$mag$]$ & 
 -- & -- & -- & -- \\
\hline
 251.73407 & -45.86723 &     SofI       & 5.4  & -- &15.51 &14.17 &15.57 &13.70 &     --        &         --        &                  --           &      --            \\
 251.76829 & -45.87432 &  PSF-rep &37.5 &37.4 & 9.84 & 8.47 & -- & -- &      W238 &     B1Iab &                   238 &    O9.5Ia-B0.5Ia \\
251.76228 &-45.83876 & PSF-rep &38.0 &37.0 & 9.58 & 8.44 & -- & -- &        W5 &   B0.5Ia+ &     5,WR77f(S) & WN10-11h/B0-1Ia+ \\
 251.75896 & -45.85353 &    2MASS   &35.7 &31.3 & -- & -- &10.30 & 8.60 &       W24 &     O9Iab &               --           &     --             \\
 251.76829 & -45.84644 &      Neg      & -- &35.5 & -- & -- & -- & -- &         W34 &      B0Ia &                 --            &      --            \\
251.78479 & -45.84592 &      Bon      & -- &26.1 & -- &-- & -- & -- &          --        &           --     &   72,WR77sc(A) &           WN7b,X \\
\hline
\footnotetext{\\
Note for the arXiv preprint version: the exact location of the online table will be finalized with the publisher. Contact the author for earlier data requests.}
\end{tabular}
\caption{Selected rows and columns from the full photometric catalog (available online). The online catalog additionally contains for each star (when available): 1) A flag indicating whether it has been kept in the analysis after $3-\sigma$ clipping, 2) A flag indicating whether it has been removed from the catalog using statistical field subtraction, 3) Its membership probability, 4) Photometric errors, 5) Optical photometry$^{1,2}$}
\label{tab:cat}
\renewcommand\tabcolsep{6pt}
\end{minipage}
\end{table*}

\subsection{Saturated stars}

The brightest stars that are present in Wd~1 can easily saturate the SofI detector.
In the following we will use the term saturated to indicate both a) stars with fluxes implying counts above the detector's full-well capacity and b) stars that are  bright enough to enter the non-linear response regime of the detector even without literally saturating it. 
The adopted SofI observations have a non-linear regime limit corresponding to $\sim 9.1$ mag in $K_{\mathrm{S}}$ band (see B08).

\texttt{DAOPHOT} cannot directly deal with saturated stars, therefore their magnitudes cannot be estimated with the same techniques adopted for the non saturated ones.
For B08 and G11 saturation of the brightest stars was not a problem. Since in those papers the authors were dealing with the intermediate and low mass content of Wd~1, well below the saturation limit, they could choose to neglect the very bright members.
However, in order to properly quantify the degree of mass segregation of Wd~1 throughout the largest possible mass spectrum, it is crucial to include the brightest and most massive objects in the present analysis. This has been accomplished by expanding the catalog of sources with near-infrared magnitude measurements in two ways:

\textbf{i. 2MASS-PSC stars:} firstly, the stars from the 2MASS-PSC were included in our source list. The 2MASS-PSC catalog for the same area was positionally matched with the SofI catalog, after excluding from the latter the saturated sources. For the stars in common we adopted the magnitudes obtained from the SofI images, given the better spatial resolution. 
We removed from the 2MASS-PSC all the sources for which a `0' read flag was present in any of the $JHK_{\rmn{S}}$ bands. 
A total of 35 sources was added to the catalog in this way. After applying the photometric selection of Sect. \ref{sec:catsel}, the number of 2MASS-PSC sources considered as members is reduced to 28. Of these, 25 are brighter than the SofI saturation limit of $K_{\rmn{S}} = 9.1$ mag, corresponding to masses larger than $\sim 30 \,\Msun$. 
The remaining 3 stars are considered as point sources in the 2MASS-PSC. However, from visual inspection, we determined that these 2MASS detections correspond to small SofI `aggregates' of 3 or 4 detections. This is a result of the poorer 2MASS angular resolution of $\sim 2\arcsec$ in the Galactic Plane, compared to a typical full width at half maximum of our SofI images of about $0\farcs75$.
Because of this ambiguity in the identification, our matching algorithm is not able to assign these 3 2MASS detections to any of its nearby SofI detections. Therefore they are considered as ``isolated'' 2MASS objects without SofI counterparts. 
Given their small number, given that they span a large range of magnitudes (and therefore masses) and also given that for increasing magnitude the number of SofI detections of similar magnitude rapidly increases, these possibly spuriously isolated 2MASS sources do not affect the outcome of our analysis, and therefore they were left in our catalog.

\textbf{ii. Repaired-PSF stars:} a second group of objects was added to our original SofI catalog. These are stars that have entered the non-linear regime of the SofI camera in either the $J$, $H$ or $K_{\rmn{S}}$ band, but for which photometry from the SofI images could still be attempted using the unsaturated wings of their light profiles. 
In order to do so we used our own adaptation of the \texttt{starfinder} code \citep{2000A&AS..147..335D}, obtaining photometry in all three $J, H$ or $K_{\rmn{S}}$ bands for 50 additional stars. 
Magnitude zero points for the repaired stars were obtained by matching their catalog with the 2MASS-PSC catalog. The total number of repaired stars also found in the 2MASS-PSC is 23. For these matched stars we adopted 2MASS-PSC magnitudes. The photometric error of the 2MASS-PSC for bright sources is indeed smaller than the precision that can be achieved by using only the PSF wings. The remaining 27 stars were added to our final catalog of detections. From the latter only 10 are kept in the list of Wd~1 members, after our photometric selection described in Sect. \ref{sec:catsel}. The remaining 17 display near-infrared colours which indicate that they are fore- or background stars.
All the objects that were added to our original SofI catalog are displayed in Fig.~\ref{fig:spdistr}, overplotted on the SofI $K_{\rmn{S}}$ band image. The repaired-PSF stars are green circles. The 2MASS-PSC sources are in turn divided in 2 groups: $K_{\rmn{S}} \leq 9.1$ mag (red crosses) and $K_{\rmn{S}} > 9.1$ mag (blue squares).

\subsection{Stars without near-infrared photometry}
\label{sec:optst}

In order to make our catalog as complete as possible, we matched the list of all the stars detected in the near-infrared with the catalogs of \cite{2007AJ....133.2696B} and \cite{Negueruela:2010fj}. The first study identifies massive stars in the Wd~1 field through photometric variability in optical bands. The second is the most complete up-to-date spectroscopic study of massive stars in Wd~1.
We will refer to the stars in these catalogs as "optical" sources in contrast to the near-infrared ones.
Given the way these two catalogs are built, we consider all their stars as members of Wd~1, therefore no additional selection was imposed to the lists of optical detections.
We matched the catalogs of optical sources with the full list of near-infrared ones, obtaining an overlap of 63 stars in total within the SofI field of view.
Among the whole set of optical sources, 10 do not have a measurement of their near-infrared magnitudes, 4 from \cite{2007AJ....133.2696B} and 6 from \cite{Negueruela:2010fj}. These stars are marked in Fig.~\ref{fig:spdistr}.

\section{Catalog selection}
\label{sec:catsel}

We adopted the same statistical membership selection scheme of G11, which is based on measuring the density of stars in the $J$ vs. $K_{\rmn{S}}$ magnitude-magnitude space. The density in one point is computed 
by summing up the contributions of all stars. These contributions are computed as Gaussians centered on the individual measurements, with corresponding covariance matrices (a case of variable, 2D, kernel density estimation). 
The densities for the science frame and the comparison field are computed separately and the contrast between the two is used as a measure of membership probability.

Once that the membership probability is computed, the catalog selection is done as follows:
for each star a uniform random number between 0 and 1 is drawn and compared with the membership probability of that star. If the former is larger than the latter, the star is rejected.
This way slightly different catalogs are obtained for different drawings. However the whole Wd~1 main sequence is easily distinguishable from the fore- and background contaminants and most of the stars with $\textrm{M} \gtrsim 3 \,\Msun$ have large membership probabilities. Therefore the different catalogs are very similar in this mass regime and variations are observed only below $\textrm{M} \sim 3 \,\Msun$. We have nevertheless verified that there are no significant variations of our results from one random realization of the catalog to the other, therefore from now on we will limit our analysis to only one particular realization.
Furthermore, in our analysis we limit ourselves to stars more massive than $3.5\,\Msun$, because this is the mass at which the average completeness for the SofI observations drops to 50\% (G11). 

\begin{figure}
 \centering
	\resizebox{\hsize}{!}{\includegraphics{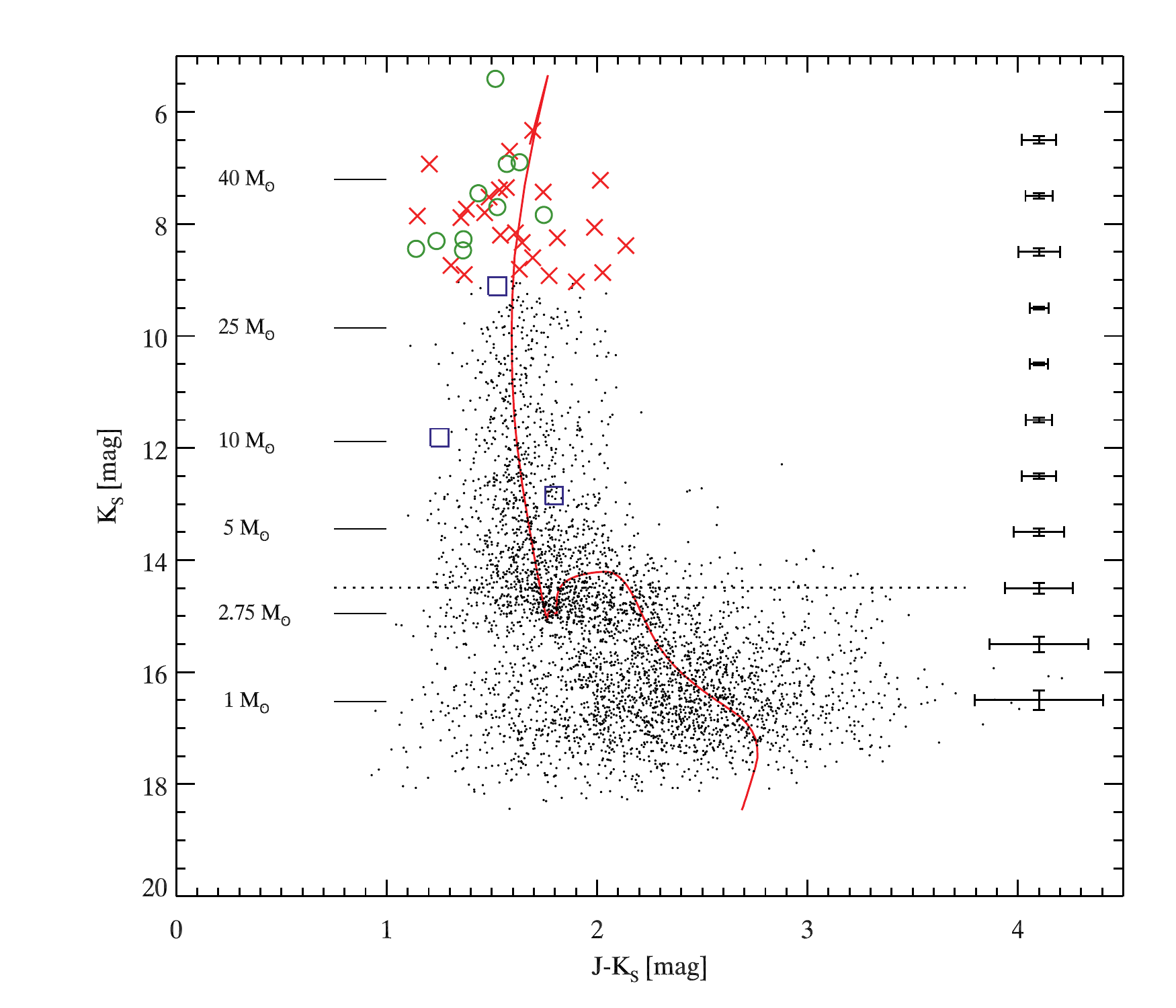}}
	\caption{Color magnitude diagram for the stars detected in the near-infrared after statistical field subtraction and 3-$\sigma$ clipping: SofI detections (black dots), 2MASS sources brighter than $K_{\rmn{S}} = 9.1$ (red crosses), the other isolated 2MASS sources with $K_{\rmn{S}} \in [10.4,14.2]$ mag (blue squares) and sources for which magnitudes were derived by PSF reconstruction through wings fitting (green circles). The red solid line is the 4 Myr isochrone described in Sect.~\ref{sec:models}. Some mass values are indicated for reference. The dotted line indicates the magnitude limit corresponding to 50\% completeness, equivalent to the main-sequence mass limit of $3.5 \,\Msun$ adopted for mass segregation estimates.}
 \label{fig:CMDcl}
\end{figure}

In addition to the probabilistic subtraction, we apply a 3-$\sigma$ clipping to the catalog to exclude all the stars whose photometry is inconsistent with the adopted isochrone of Sect.~\ref{sec:models} which traces the bulk of Wd~1 population (see again G11). The number of objects from each source catalog before and after the clipping and statistical subtraction steps are summarized in Table \ref{tab:Catcounts}; the colour magnitude diagram for the 3997 selected members is shown in Fig.~\ref{fig:CMDcl}.

\begin{table}
\centering
\begin{tabular}{lccc}
\hline
\hline
Original & Total & After   & After\\
catalog & detections & clipping & subtraction \\
\hline
SofI & 5994 & 4638 & 3949\\
Repaired-PSF & 42 & 28 & 28\\
2MASS & 27 & 10  &  10\\
Bonanos 2007 & 8 & 4 & 4\\
Negueruela 2009 & 9 & 6 & 6\\
\hline
Total & 6080 & 4686 & 3997 \\
\hline
\end{tabular}
\caption{\label{tab:Catcounts}Number of sources from each catalog.}
\end{table}

\section{Stellar models}
\label{sec:models}

For the determination of stellar masses, we compared our photometry with stellar models. 
The adopted isochrone is the 4 Myr, solar composition isochrone described in G11. It is a match of an isochrone from \cite{Marigo:2008lr}, well suited for describing the MS of Wd~1 and an isochrone from \cite{DeglInnocenti:2008fk}, better suited for the stars that are still in the pre-main-sequence (PMS) phase.
The match between the two isochrones is done at $4 \,\Msun$. According to these models, a $\sim 2.75\, \Msun$ star is just on the zero-age-main-sequence position, with more massive stars in the MS and less massive still in the PMS.
The turn-off mass, i.e. the mass for which central hydrogen exhaustion occurs, is approximately $40 \,\Msun$. After hydrogen exhaustion stars rapidly expand and strongly increase their luminosity during their short lived post-MS phase.

The age and distance adopted are the same as in G11, i.e. $t = 4$ Myr and DM = 13.0 mag, corresponding to 4.0 kpc.
The reddening value was derived by fitting the vertical upper-MS part of the isochrone to the stars in the magnitude interval $K_{\rmn{S}} \in [9.1,13.5]$ and colour $J-K_{\rmn{S}} \in [1.2, 2.0]$.
We obtained the reddening value by minimizing the quantity: 
\[
\sum_j | (J - K_{\rmn{S}})_j - (J - K_{\rmn{S}})_{isoc}| 
 \]
where j runs over the selected stars and the isochrone colour is taken at the same $K_{\rmn{S}}$ of the j-th star.
Once the $J - K_{\rmn{S}}$ reddening has been estimated, extinction $A_{K_{\rmn{S}}}$ is computed using the extinction law by \cite{2006ApJ...638..839N}. 
Given the slightly different photometric calibration, the derived value of $A_{K_{\rmn{S}}}$ changes from 0.907 mag of G11 to 0.913 mag in the present work, a negligible difference.

\section{Individual masses}
\label{sec:indmass}

The stellar mass for the objects with near-infrared photometry was determined with a maximum-likelihood method, by comparing the measured magnitudes with the isochrone of Sect.~\ref{sec:models}. 
For the i-th star we assumed Gaussian photometric errors and computed the probability of the photometric data, given the mass (i.e. the likelihood):

\begin{eqnarray}
\label{eq:lkl}
\lefteqn{ p(\mathbf{d}_i|m) = \frac{1}{2 \upi |\mathbf\Sigma_i|^{1/2}} \times } \nonumber \\
\lefteqn{\qquad \exp \left\{ -\frac{1}{2}  \left[\mathbf{M}(m) - \mathbf{d}_i\right]^T {\mathbf\Sigma_i}^{-1} \left[\mathbf{M}(m) - \mathbf{d}_i\right] \right\} \; ; }
\end{eqnarray}
where $\mathbf{d}_i = (J_i, K_{\rmn{S}\,i})$, with $\mathbf\Sigma_i$ being the covariance matrix for the individual star (magnitude errors and their covariance), and $|\mathbf\Sigma_i|$ its determinant. The matrix is derived as described in appendix B of G11.
$\mathbf{M}(m) = (J(m), K_{\rmn{S}}(m))$ are the values of the magnitudes predicted by our chosen model, for stellar mass $m$.

The probability distribution function (pdf) of the mass given the data, $p(m|d)$, is proportional  to the product of $p(d|m)$ and $p(m)$, the latter being the probability of the model (prior).
In G11 the authors found that, for the range of masses of interest here, the width of the likelihood (as a function of mass) for individual stars was of a few percent of the peak mass or less.
Given that, for simplicity, we adopt here a flat prior for $m$ throughout the observed mass range; therefore the mass pdf is proportional (though with different units) to the likelihood.
We use the mode of the likelihood as our mass estimate for the individual stars.

The sorted list of stellar masses is what really matters for our analysis, while the exact mass values are not of fundamental importance. We do not expect any systematic uncertainty in our results to arise from the choice of the prior.

From Fig.~\ref{fig:CMDcl} it is possible to see that the 50\% completeness magnitude (dotted line), which we adopt as a cut-off for our analysis, corresponds to a region where the isochrone assumes the same magnitude values multiple times. This degeneracy corresponds to the PMS to MS transition, where the lower-mass PMS stars that are about to enter the zero-age MS are actually brighter than slightly more massive stars already in MS.
The likelihood function for stars observed in this region can thus show multiple peaks and so the mode, which we adopt as a mass estimator, will correspond to point of the isochrone which is closer to the measured magnitudes \citep[see][for a detailed description]{2012MNRAS.420..986G}. Nevertheless, this mass ambiguity affects only the lowest mass stars and it is not important for our general analysis and conclusions, which are mostly focused on the behaviour of the massive stars.

\subsection{Mass of the optical sources}

For the 10 sources without near-infrared detections described in Sect.~\ref{sec:optst} we estimated the mass in a similar way as for the ones with near-infrared detections, but using $R$ and $I$ magnitudes.
Since these stars are all upper-MS objects, it was sufficient to use the 4 Myr MS isochrone by \cite{Marigo:2008lr}, provided in the appropriate photometric filter system.
Given the $A_{K_{\rmn{S}}} = 0.913$ mag, estimated from the near-infrared observations, we calculated the corresponding $A_R$ and $A_I$ values by adopting $A_J/A_{K_{\rmn{S}}} = 3.02$ value from the \cite{2006ApJ...638..839N} reddening law, which is determined only down to $J$ band and extended it to shorter wavelengths using the $A_R/A_J = 2.66$ and $A_I/A_J =1.70$ by \cite{1989ApJ...345..245C}.

\section{Structure and mass segregation in Wd~1}
\label{sec:structure}

We quantify the structure and mass segregation in Wd~1 in three ways:
using the $\mathcal{Q}$-parameter \citep{2004MNRAS.348..589C}, the mass segregation ratio 
$\Lambda_{\rm MSR}$ \citep{allison09}, and the local surface density ratio for massive stars $\Sigma_{\rm LDR}$
\citep{2011MNRAS.416..541M, 2014MNRAS.438..620P}.  We will quickly summarise
these methods below, but we direct the reader to \cite{2014MNRAS.438..620P} and
\cite{2015MNRAS.449.3381P}
for a detailed discussion of these
methods, their pros and cons, and variations on them.

The $\mathcal{Q}$-parameter takes the ratio of the average minimum spanning tree
(MST) path length, $\bar{m}$, to the average distance between stars, $\bar{s}$, to quantify
the level of substructure:
\begin{equation}
\mathcal{Q} = \frac{\bar{m}}{\bar{s}}.
\end{equation}
If $\mathcal{Q}<0.7$ there is significant
`clumpiness', if $\mathcal{Q}>0.9$ the structure is smooth with increasing
central concentration with increasing values of $\mathcal{Q}$.  (Essentially,
clumpy regions will contain many more short connections.)

The $\Lambda_{\rm MSR}$ mass segregation ratio compares the MST length connecting the $N_{\rm MST}$ most
massive stars (or any subset of stars), $l_{\rm subset}$, with the average MST length of
large ensembles of sets of $N_{\rm MST}$ random stars, $\langle l_{\rm average} \rangle$.
There is a dispersion  associated with the average length of
random MSTs, which is roughly Gaussian and can be quantified as the
standard deviation  of the lengths  $\langle l_{\rm average} \rangle
\pm \sigma_{\rm average}$. However, we conservatively estimate the lower (upper) uncertainty 
as the MST length which lies 1/6 (5/6) of the way through an ordered list of all the random lengths (corresponding to a 66 per cent deviation from 
the median value, $\langle l_{\rm average} \rangle$). This determination 
prevents a single outlying object from heavily influencing the uncertainty.
$\Lambda_{\rm MSR}$ is thus given by
\begin{equation}
\Lambda_{\rm MSR} = {\frac{\langle l_{\rm average} \rangle}{l_{\rm subset}}} ^{+ {\sigma_{\rm 5/6}}/{l_{\rm subset}}}_{- {\sigma_{\rm 1/6}}/{l_{\rm subset}}},
\end{equation}
and quantifies the extent
to which the MST length of the most massive stars is likely to be
randomly sampled. In this work we utilize 100 random ensembles to estimate the uncertainty on $\Lambda_{\rm MSR}$.
  If $\Lambda_{\rm MSR} > 1$ the most massive stars are
concentrated (mass segregated), if $\Lambda_{\rm MSR} < 1$ the massive stars are
more widely dispersed than expected at random.

\cite{2011MNRAS.416..541M} quantified the relative spatial distribution of massive stars compared to the cluster average using the $\Sigma - m$ plot, which plots the local surface density, $\Sigma$ around each star as a function of its mass, $m$. $\Sigma$ is given by 
\begin{equation}
\Sigma = \frac{N - 1} {\pi r_{N}^2},
\end{equation}
where $r_{N}$ is the distance to the $N^{\rm th}$ nearest neighbouring star (we adopt $N = 10$ throughout this work). The median density of the chosen subset of stars is then compared to the median density for the entire distribution \citep{2011MNRAS.417.2300K,2014MNRAS.438..620P} to obtain the local surface density ratio, $\Sigma_{\rm LDR}$:
\begin{equation}
\Sigma_{\rm LDR} = \frac{\tilde{\Sigma}_\mathrm{subset}}{\tilde{\Sigma}_\mathrm{all}}
\end{equation} 
The significance of this measure of the local density of a subset of stars compared to the cluster is defined by a Kolmogorov-Smirnov (KS) test between the $\Sigma$ values of the subset against the $\Sigma$ values of the rest.

The $\Sigma_{\rm LDR}$ measure compares the local surface densities of
the most massive stars with the average local surface density and
examines the probability that the most massive stars are in regions of
higher surface density than would be expected at random.

Note that $\Lambda_{\rm MSR}$ and $\Sigma_{\rm LDR}$ will both find `mass
segregation' when the most massive stars are at the centre of a single
cluster.  However, they measure different types of `mass segregation':
the relative positions of the massive stars for $\Lambda_{\rm MSR}$, and the
local surface densities for $\Sigma_{\rm LDR}$ and so it is quite
possible to have a `signal' in one but not the other method \citep[see][]{2015MNRAS.449.3381P}.

Both methods rely on the determination of individual stellar masses for all cluster members. In a future paper we will study the full effects of observational uncertainties on $\Lambda_{\rm MSR}$ and $\Sigma_{\rm LDR}$; however, we note from our previous work that uncertainties on individual stellar masses of up to 30\,per cent would not alter a strong signature of mass segregation \citep{2011MNRAS.412.2489P}.

We examine two samples from the data, one in which we assume
completeness above 5~\Msun, and one above 3.5~\Msun.  We find
little difference between these two samples.

\subsection{Substructure in Wd~1}

We calculate $\mathcal{Q}$ for our two samples, and find $\mathcal{Q}=0.90$ for all
stars above 5~\Msun, and $\mathcal{Q}=0.87$ for all stars above
3.5~\Msun.  These $\mathcal{Q}$-values are those of a fairly smooth, 
somewhat centrally concentrated cluster which matches what 
the eye sees in Fig.~1.  [Note that elongation can bias $\mathcal{Q}$ somewhat to lower
values than would be expected \citep{2009MNRAS.392..868B}.]

A value of $\mathcal{Q} \sim 0.9$ for Wd~1 is interesting.  There is no evidence
for any significant substructure or subclustering (which would give $\mathcal{Q}
< 0.7$).  But there is also no evidence for a significant central
concentration.  A $\mathcal{Q} \sim 0.9$ would suggest a density distribution
with a power-law of $r^{-2}$ \citep[possibly slightly shallower,][]{2004MNRAS.348..589C}. 

It is unclear what biases there could be in the determination of $\mathcal{Q}$.
If significant numbers of low-mass stars are missed in the central
regions, then $\mathcal{Q}$ would be lowered \citep[this would also {\em lower} both
$\Lambda_{\rm MSR}$ and $\Sigma_{\rm LDR}$, see][]{2012MNRAS.426.3079P}.   Missing low-mass stars near
high-mass stars would probably lower $\mathcal{Q}$ somewhat as well by producing
artificial `holes' around massive stars (this would artificially raise
$\Lambda_{\rm MSR}$ and lower $\Sigma_{\rm LDR}$).  Therefore we suggest that the measured
$\mathcal{Q} \sim 0.9$ is probably a lower limit on the true $\mathcal{Q}$. 

\subsection{Mass segregation in Wd~1}

\begin{figure}
  \begin{center}
  \includegraphics[width=0.5\textwidth]{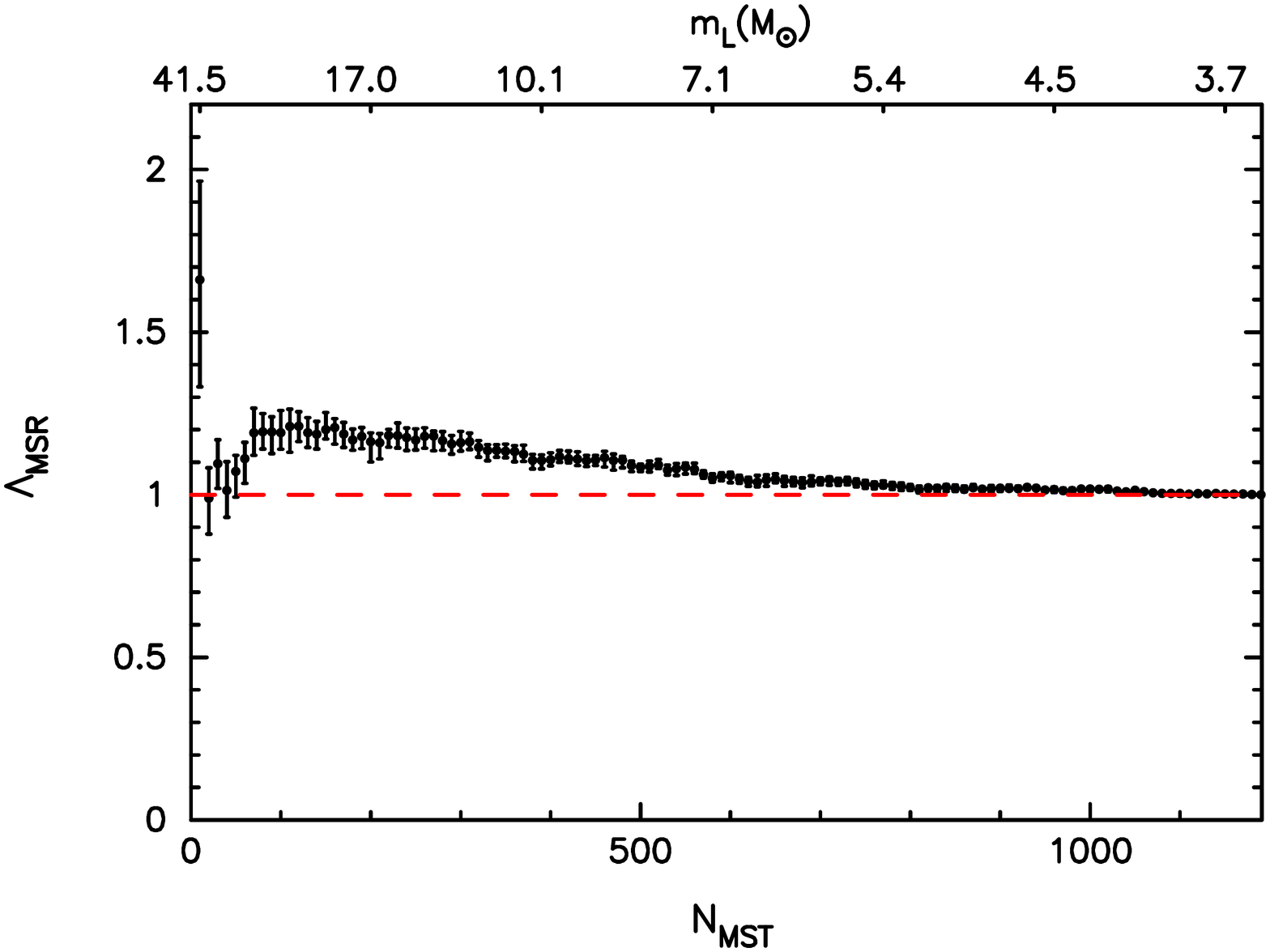}
  \includegraphics[width=0.5\textwidth]{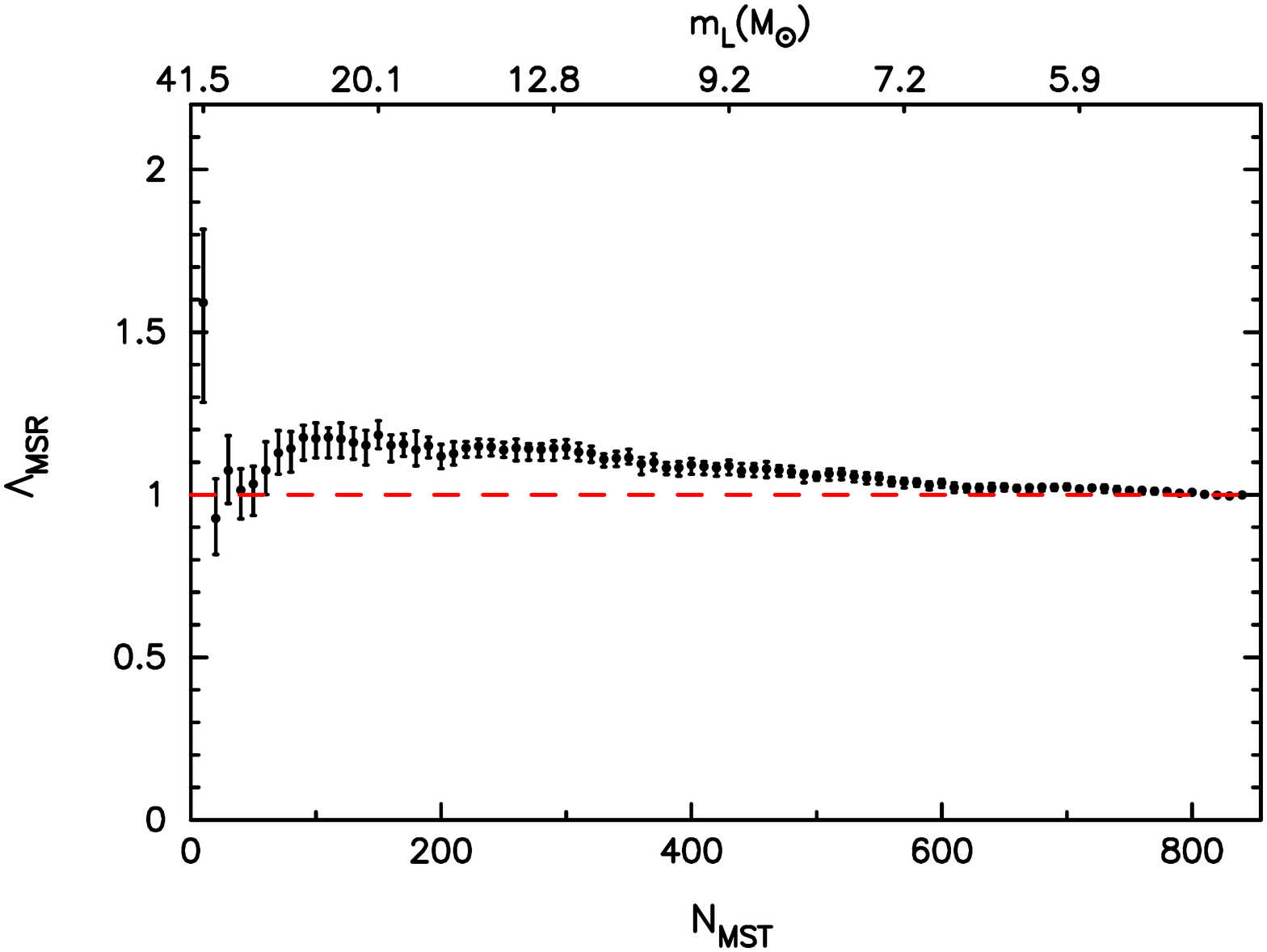}
  \caption[]{The evolution of $\Lambda_{\rm MSR}$ for increasing values of $N_{\rm MST}$. The red dashed line indicates $\Lambda_{\rm MSR}$  = 1, i.e. no mass segregation. The top axis indicates the minimum stellar mass within the $N_{\rm MST}$ subset of the most massive stars.
The two panels show computations considering all the stars with M$>3.5$~\Msun (top) and all the stars with M$>5$~\Msun (bottom).\label{fig:lmst}}
  \end{center}
\end{figure}

In Fig.~\ref{fig:lmst} we show $\Lambda_{\rm MSR}$ as a function of $N_{\rm MST}$ for the $>5$~\Msun~and
$>3.5$~\Msun~samples.  The values for $\Lambda_{\rm MSR}$ are very similar for both mass cuts. $\Lambda_{\rm MSR} = 1.6^{+0.2}_{-0.3}$ for the 10 most massive
stars ($> 40$~\Msun) but is consistent with unity for lower masses. 

This is, at best, marginal evidence for mass segregation, and then
only in the most massive stars. Likewise, the dip in $\Lambda_{\rm MSR}$ which appears below 41.5 \Msun cannot be considered a significant feature. \cite{2015MNRAS.449.3381P} show that any value of $\Lambda_{\rm MSR}$ below 2 should be treated with caution, especially for low values of the $N_{\rm MST}$ stars used to construct it. In particular, they show that a random distribution of points can sometimes give an apparent signature of mass segregation. If a region is truly mass segregated \citep[e.g. the Orion Nebula Cluster][]{allison09}, $\Lambda_{\rm MSR}$ takes values very much larger than 2.

Any biases in the observations will
almost certainly increase the apparent significance of any mass
segregation as they will involve missing low-mass stars near high-mass
stars and hence artificially increasing the length of random MSTs \cite[see][]{2009Ap&SS.324..113A}.  Therefore, this value of $\Lambda_{\rm MSR}$ should be
taken as an upper limit on the true value. 
\subsection{Local surface densities of massive stars in Wd~1}

\begin{figure}
  \begin{center}
  \includegraphics[width=0.5\textwidth]{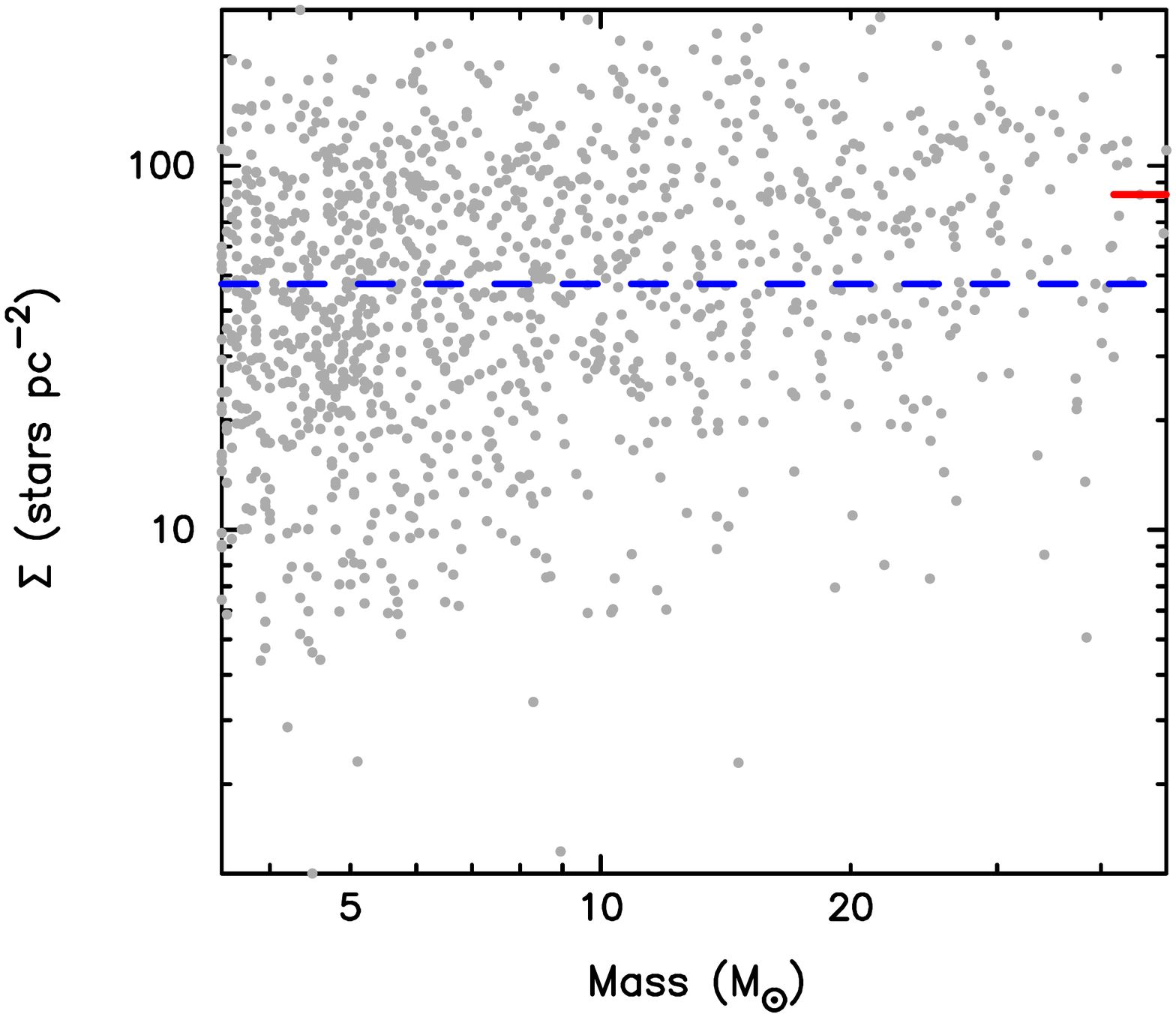}
  \includegraphics[width=0.5\textwidth]{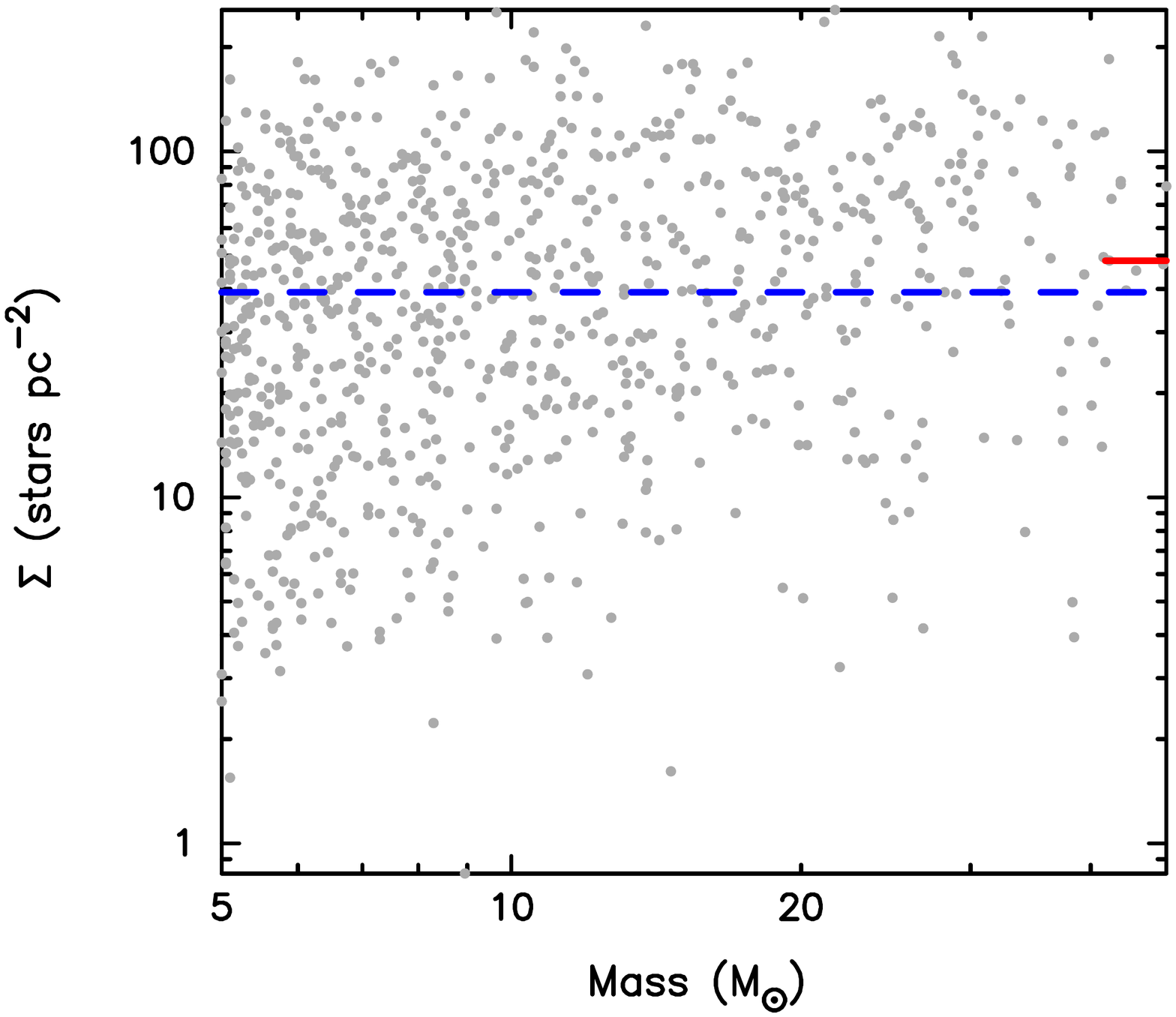}
  \caption[]{The stellar surface density, $\Sigma$, as a function of stellar mass. The blue dashed line indicates the median surface density for the full sample, and the solid red line indicates the median surface density for the most massive stars. The top panel shows all stars with M$>3.5$~\Msun, the bottom all stars with M$>5$~\Msun\label{fig:sigma}.}
  \end{center}
\end{figure}

In Fig.~\ref{fig:sigma} we show the plots of surface density as a function of stellar mass to determine
$\Sigma_{\rm LDR}$ for both the $>5$~\Msun and $>3.5$~\Msun
~samples. The panels show the surface densities for all stars for the
$>3.5$~\Msun ~sample (top panel), and $>5$~\Msun~ sample (bottom
panel).  In both panels the mean surface density is shown by the blue, dashed
line, and the mean of the surface densities for the 10 most massive
stars by the red line.

In both cases we find that the local surface density around
the most massive stars is higher than that around low-mass stars.  In
the $>5$~\Msun~sample the difference is a factor of 1.23 which has
a Kolmogorov-Smirnov (KS) test p-value of 0.06; in the $>3.5$~\Msun~sample the difference is a
factor of 1.76 with a p-value of 0.04.    

The raw values of $\Sigma_{\rm LDR}$ are different in the two samples
because the total number of stars in each sample is different which
results in a higher average surface density in the $>3.5$~\Msun
~sample (top panel) so the raw values should be treated with some caution.   

However, both samples show that the most massive stars in Wd~1 are
located in regions of significantly higher-than-average surface density.   

For $\Sigma_{\rm LDR}$ the effect of incompleteness near massive stars
would be to reduce a signal (i.e. missing low-mass stars near massive
stars would artificially lower their local surface density).  Therefore
we conclude that the massive stars in Wd~1 are almost certainly found
in regions of higher than average surface density. 

\subsection{Comparison with previous work}

It should be noted that the result of little or no mass
segregation in Wd~1 (and if any, only above 40~\Msun) condradicts that
of G11.  
The difference is due to the method used in G11 in
fitting IMF slopes in different areas which involves two sets of
somewhat arbritary `binning' (see their section 8.3).  We 
suggest that the problems with binning and fitting slopes 
(especially at the high-mass end where there are few stars) resulted
in an over-estimate of the degree of mass segregation.  

Parker \& Goodwin (2015) discuss the problems of determining mass
segregation.  They conclude that $\Lambda_{\rm MSR}$ is the most reliable method
of determining `classical' mass segregation.  In particular, it does
not involve determining an IMF slope with all of its associated 
problems, rather it simply tests the hypothesis that the most massive
stars are compatible with being a random subset of all stars.  As
discussed above, {\em all} methods are biased by possible biases in the data
such as missing low-mass stars near high-mass stars.
However $\Lambda_{\rm MSR}$ is biased in such a way that it is an upper limit on
the true $\Lambda_{\rm MSR}$, whilst IMF fitting would be biased such that it
would find a flatter slope and hence `find' mass segregation.

\subsection{Summary of observational results}

The spatial analysis of Wd~1 can be summarised as follows.

Wd~1 shows little evidence of substructure with a $\mathcal{Q}$-value of $\sim
0.9$ which is slightly centrally concentrated (a density profile of
roughly $\sim r^{-2}$).   

Wd~1 shows only marginal evidence for stars $>40$~\Msun~ being a factor
of $\sim 1.5$ times more concentrated than random, and no evidence of
any other stars being mass segregated at all. 

Wd~1 shows good evidence (KS test P-values of $\sim 0.05$) of the most
massive stars being in regions of higher local surface density than
other stars.   

The most probable bias would be to miss low-mass stars near high-mass
stars.  The result of such a bias would be that the measured value of
$\mathcal{Q}$ is a lower limit, of $\Lambda_{\rm MSR}$ an upper limit, and of $\Sigma_{\rm LDR}$ a
lower limit. 

\section{Discussion}
\label{sec:discuss}

The results of our analysis of the structure of Wd~1 and the relative
distributions of its high- and low-mass stars enable us to speculate
on the initial conditions and evolution of Wd~1.

\cite{2014MNRAS.438..620P} examine the evolution of $\mathcal{Q}$, $\Lambda_{\rm MSR}$ and
$\Sigma_{\rm LDR}$ in a selection of $N$-body simulations with a variety
of different initial conditions (clumpy, hot, cool, smooth etc.).  
They find the following general conclusions.

Firstly, $\mathcal{Q}$ either stays the same or increases with time.  In unbound
regions the initial structure (and hence the $\mathcal{Q}$-parameter) is `frozen-in' and
does not change with time.  In bound regions any substructure is
erased and $\mathcal{Q}$ increases.  Therefore the current $\mathcal{Q}$ is an upper limit
on the initial $\mathcal{Q}$.

Secondly, $\Sigma_{\rm LDR}$ increases with time.  Massive stars are
able to gain a `retinue' of low-mass stars over time which increases their
local surface density.  Therefore the current $\Sigma_{\rm LDR}$ is an 
upper limit on the initial $\Sigma_{\rm LDR}$.

Finally, $\Lambda_{\rm MSR}$ can evolve in a number of ways.  The most massive
stars can dynamically mass segregate as a bound cluster collapses \citep[see][]{2009ApJ...700L..99A}, but a mass segregated cluster can also be
dynamically destroyed by multiple instability or binary formation
leading to the ejection of the most massive stars \citep[see also][]{2010MNRAS.407.1098A,2011MNRAS.415.1967A,2014MNRAS.438..620P}.  Therefore the current value of $\Lambda_{\rm MSR}$ must be
interpreted carefully. 

We can also include additional information on the dynamics of Wd~1
which suggests the region is close to having a virial velocity
dispersion\footnote{This is that the velocity dispersion is close to
  that expected for a region in virial equilibrium, although we note
  that it is probably not in statistical equilibrium.}, and is possibly
sub-virial \citep{cottaar12}.

\subsection{The formation and evolution of Wd~1}

What were the initial conditions of Wd~1?

Firstly, Wd~1 may have formed with or without any primordial mass segregation.

Secondly, in its initial size there are three general scenarios that we can imagine for Wd~1:\\ 
1. {\bf Expanded.} Wd~1 was initially much more dense than we see now and has expanded to its current size (initial size $< 1$~pc).\\ 
2. {\bf Static.} Wd~1's initial
conditions were very similar to what we see now (initial size $\sim 2$~pc).\\ 
3. {\bf Collapsing.} Wd~1 was initially larger and
dynamically cool, and is collapsing (initial size $> 3$~pc).

Note that there is also a possible `hybrid' scenario:\\
4. {\bf Collapse and bounce.}  Wd~1 formed at a similar or larger size than we see now, collapsed to a denser state than we see now and has `bounced' to its current size (e.g. Allison et al. 2009; Allison \& Goodwin 2011).

\bigskip

Firstly, it is worth considering what mechanisms would be responsible for
  the behaviour in each of the scenarios.

Wd~1 will presumably have formed from a GMC significantly more massive than
the mass of stars we now see.  It is impossible to guess the star
formation efficiency of the GMC that formed Wd~1, but a value of
around 10 per cent seems reasonable (with potentially very significant
local-to-global variations).

In the expanded scenario, the expulsion of the gas remaining after star
formation would presumably be the driver of the expansion
\citep[e.g.][]{2006MNRAS.373..752G,2007MNRAS.380.1589B}, although it is worth
noting that stellar dynamics alone can cause significant expansion of
clusters  \citep{2012MNRAS.426L..11G,2015arXiv151004293B,2012MNRAS.425..450M,2012MNRAS.427..637P,2013A&A...559A..38P,2013MNRAS.432..986P}.

In the collapsing scenario, Wd~1 must have been relatively cool (low
virial ratio) otherwise it would not be able to collapse.  In this
case the effect of gas loss is nowhere near as dramatic as for that
from a relaxed cluster due
to the strong dependence of the effect of gas expulsion on the global
virial ratio \citep[][Lee \& Goodwin 2016]{2006MNRAS.373..752G,2009Ap&SS.324..259G,2013MNRAS.428.1303S}.  

In the static case, one could imagine a situation where Wd~1 was initially cool
enough that the gas loss was balanced (this would not need to be
exact;  we would still class as `static' changes in radius between 1.5 or 2.5~pc and the current 2~pc
size).  

In the collapse and bounce case, Wd~1 forms dynamically cool (as in the collapsing scenario), collapses, reaches a very dense state, and `bounces' due to heating from dynamical mass segregation, and possibly massive binary formation (see Allison et al. 2009; Allison \& Goodwin 2011).

Note that the bulk dynamical effects of mass loss
  due to stellar evolution will be relatively small.  With a standard
  IMF a massive cluster is expected to have lost $\sim 10$ per cent of
  its mass in 5~Myr \citep{2014MNRAS.445.1967T}.  From an initially
  virialised cluster this would cause an expansion by a factor of
  $\sim 10$ per cent.  Therefore stellar evolution is irrelevant in
  our discussion of significant expansion or collapse, although we note that if the cluster was significantly mass segregated, mass loss would occur in the central region of the cluster (a region with high local binding energy). If those stars lost significant mass, the cluster could potentially expand at a faster rate than if it was not mass segregated.
  
\bigskip

It is important to note that in each of these cases the {\em dynamical} age of Wd~1 is very
different, and it is from this difference that we might hope to
disentangle the history.

The {\em current} crossing time of Wd~1 with a mass of $5 \times
10^4$~M$_\odot$ and a radius of 2~pc -- and assuming it is virialised
-- is $\sim 0.2$~Myr.  Therefore the physical age of 5~Myr is roughly
25 current crossing times.  

In the expanded and bounced scenarios Wd~1 will be (much) more
than 25 crossing times old, in the static scenario it will be roughly
25 crossing times old, and in the collapsing scenario it will be
(much?) less than 25 crossing times old.  Therefore, different scenarios produce a present-day 
cluster with very different dynamical ages. 

Dynamical age is important in erasing initial substructure, but mostly in that dynamical evolution causes some level of dynamical mass segregation.

In a cluster the timescale $t(M)$ of a star of mass $M$ to reach
energy equipartition (hence to mass segregate) is
\begin{equation}
t(M) \sim \frac{\bar{m}}{M} \frac{N}{8~{\rm ln}~N} t_{\rm cross},
\end{equation}
where $\bar{m}$ is the average mass of a star in the cluster (about
0.4~\Msun), $N$ is the total number of stars in the cluster, and
$t_{\rm cross}$ is the crossing time of the cluster. 

Depending on the {\em dynamical} age of Wd~1 it is a different number of crossing times old, and so the mass to which mass segregation can occur changes.  If we assume Wd~1 is in virial equilibrium (ie. the static case) and take $R=2$~pc and $M_{\rm cl} = 5 \times 10^4$~\Msun~
then in 5~Myr we would expect to mass segregate down to $\sim
20$~\Msun.  (This is somewhat lower than the $\sim 40$~\Msun~ down
to which we possibly see mass segregation.)

Now we consider each of the scenarios in turn.

\bigskip

{\bf Expanded or collapse/bounce scenarios.} In both of these scenarios Wd~1 would have been significantly more dense at some point in the past than it is now.  In its densest phase the crossing time would have been much shorter allowing significant dynamical evolution to have occured.

The effects of significant dynamical evolution would have been to erase any substructure and to allow dynamical mass segregation of the most massive stars to occur.

Dynamical mass segregation of the most massive stars occurs very rapidly in a dense state (e.g.  Allison et al. 2009) meaning that mass segregation would be expected even if it was not primordial (primordial mass segregation would be difficult to distinguish from dynamical mass segregation in such a system).

Therefore the initial (or very early) $\Lambda_{\rm MSR}$ should have been
significantly $>1$, i.e. much higher than we now observe it to be.  However, this mass segregation produces marginally stable `Trapezium-like' systems which sometimes decay (Allison \& Goodwin 2011).  Usually the decay of such systems ejects some of the most massive stars and creates a signature of `inverse mass segregation' (ie. $\Lambda_{\rm MSR} < 1$; see Allison \& Goodwin 2011; Parker et al. 2014), but it is possible for a mass segregated cluster to move to a state where mass segregation is minimal.  This is unlikely in such a scenario, but not so unlikely as to render the scenarios completely untenable.

The expansion must also be such that the cluster has had chance to
revirialise by its current 4--5~Myr age which is quite possible \citep[see e.g.][]{2013ApJ...764...29B} but would require a little fine-tuning. And after
the expansion there must have been sufficient time for the initially
mass segregated massive stars to accumulate a retinue of low-mass
stars to increase their local surface density to the significantly
higher than average values now seen. Recent work by \citet{2016MNRAS.457.3430P} has shown that a collapse and bounce would likely ``freeze in" an artificially high velocity dispersion, making the cluster appear to be supervirial when in fact it is in virial equilibrium (though perhaps not in statistical equilibrium). Whilst these simulations contain at least ten times fewer stars than Wd~1, the artificially high velocity dispersion is caused by the high degree of violent relaxation, which in turn is a function of the initial conditions, including a high initial stellar density, $10^4$\,M$_\odot$\,pc$^{-3}$, rather than the number of stars. Simulations with a lower initial density, and a less violent, more prolonged collapse, do not display this effect \citep{2016MNRAS.457.3430P}. If the collapse and bounce scenario had already occured for Wd~1, we would therefore expect to see a high velocity dispersion. Given that \citet{cottaar12} measure a velocity dispersion consistent with a (sub)virial state for Westerlund 1, we suggest that the collapse and bounce scenario is unlikely.

\bigskip

{\bf Static.}  It is possible that we see Wd~1 at an age of 5~Myr as
appearing quite similar to how it formed -- a 2~pc diameter, somewhat
elongated, roughly virialised, bound cluster.   

Wd~1 is, at most, mass segregated to $\sim 40$~\Msun.  From our argument above (see eqn.~6) we would expect mass segregation in the static case to a mass probably somewhat lower ($\sim$ 20\,M$_\odot$) than we actually observe it, although not significantly so given the number of assumptions we are making.

However, Wd~1 is segregated to roughly the level we would expect from an initially unsegregated cluster.  Any primordial mass segregation would be enhanced, and the only way to fail to see evidence of primordial mass segregation would be if it was such that it matched what is expected dynamically (which raises the question of how Wd~1 `knew' to primordially segregate to match the effect of dynamics).

As shown by \cite{2014MNRAS.438..620P} we would expect $\mathcal{Q}$
to rise somewhat 
from its initial value as substructure is erased.  

During evolution we would also expect the massive stars to collect significant retinues of
low-mass stars into their potential wells.  Thus we would expect
$\Lambda_{\rm MSR}$ to be slightly greater than unity, but not much
\citep[see][their Fig. 6 top row]{2014MNRAS.438..620P}, and  $\Sigma_{\rm
  LDR}$ to show a quite significant  
signature \citep[see][their Fig. 7 top row]{2014MNRAS.438..620P}. 

An elongated cluster is quite plausible in a static model as it has not had time to completely erase any primordial anisotropies (in both velocity and/or spatial structure).

\bigskip

{\bf Collapsing.}  In a collapsing model Wd~1 could have formed at
maybe roughly twice the size we see now and be collapsing into a
dense cluster;  it cannot have formed at much more than twice the size
we see now in order to have collapsed to its current state in 5~Myr.
This is a more massive version of the scenario for the formation of
the ONC proposed by \cite{2011MNRAS.415.1967A}\footnote{The
    degree to which a distribution can collapse depends on the initial
    virial ratio and the degree of substructure.  A cool, very clumpy
    distribution can collapse by a factor of 2--3 \citep{2011MNRAS.415.1967A}.}.

As mass segregation is driven by dynamics we would expect it to be
somewhat slower than the prediction for the more dense static case given above
(which resulted in a lower limit of $\sim 20$~\Msun), and we estimate that the collapsing scenario would mass
segregate Wd~1 down to a mass of $40-50$~\Msun.  This is a better fit to
the observed value to which mass segregation is seen, but we would be
hesitant to make too much of this as it involves many assumptions
about the evolution.

The same arguments apply to massive stars collecting retinues of low-mass stars, erasure of substructure, and elongation as in the static case. 

\bigskip

{\bf Summary.} With the current data it is impossible to completely rule-out any scenario for the evolution of Wd~1.  It may have expanded, be collapsing, have collapsed and bounced, or have formed simlar to the way we now see it.  However, we can make some arguments as to which are the most likely.

Some dynamical mass segregation is expected in all scenarios as the effect of relaxation will be to remove energy from the most massive stars.  In the collapsing or static scenarios dynamics would be expected to give a signature very similar to what we observe (minor mass segregation of only the most massive stars).  The massive stars have had sufficient time to collect a retinue of lower-mass stars to give them a high relative surface density, and elongation would be a memory of anisotropic/clumpy initial conditions that have not been erased as the cluster is not dynamically evolved enough to have done this.

However, in the expanded or collapse/bounce scenarios many different possible mass segregation signatures can arise (from very mass segregated to inverse mass segregation), so they 
require a little `luck' in finishing in a state in which any primordial/dynamical mass segregation is erased such that $\Lambda_{\rm MSR} \sim 1$ at the end.  If the cluster was lucky enough to expand with $\Lambda_{\rm MSR} \sim 1$ then the massive stars would have had sufficient time to collect a retinue of lower-mass stars.

It is very worth noting that in both the static and collapsing scenarios the mass segregation is exactly what would be expected from dynamics alone, and therefore suggests that Wd~1 was {\em not} primordially mass segregated.  In the expanded or collapse/bounce scenarios primordial mass segregation may have been present, but significant dynamical segregation would have occurred, and then have been erased.

It is unclear whether the absence of primordial mass segregation has
any implications for theories of star formation. Primordial mass
segregation appeared to be ubiquitous in early simulations advocating
competitive accretion as the dominant channel of massive star
formation \citep{1997MNRAS.285..201B,2001MNRAS.323..785B}. However, in
recent simulations that include more diverse physical processes such
as feedback, primordial mass segregation is not present
\citep{2015MNRAS.446.4278P,2017MNRAS.470..390P}.

\section{Summary and Conclusion}
\label{sec:conc}

At $\sim 5 \times 10^4$~\Msun~ and 4--5~Myr old Wd~1 is the most massive young cluster in the Galaxy. We have used a combination of near infrared photometry from NTT/SofI \citep{Brandner:2008lr} and the 2MASS Point Source Catalog \citep{Skrutskie:2006uq} as well as optical photometry \citep{2007AJ....133.2696B,Negueruela:2010fj}
to analyse the distribution of stars, and massive stars in particular, in Wd~1.

We analyse the distribution of stars to look for substructure using the $\mathcal{Q}$-parameter \citep{2004MNRAS.348..589C}.  We also look at the relative distributions of massive stars with the $\Lambda_{\rm MSR}$ mass segregation ratio \citep{allison09}, and also the relative local surface densities of massive stars with the $\Sigma_{\rm LDR}$ local surface density ratio \citep{2011MNRAS.416..541M,2014MNRAS.438..620P}.  We find that Wd~1:\\

\noindent (i) Is relatively smooth, and slightly centrally concentrated with $\mathcal{Q} \sim 0.9$.\\
(ii) Shows little evidence for mass segregation with $\Lambda_{\rm MSR} = 1.6$, but even then only for stars with masses $>40$~\Msun.\\
(iii) Has massive stars preferentially in regions of higher than average
surface density. \\

It is also known that Wd~1 is at most virialised, and quite possibly
sub-virial \citep{cottaar12}. 

We critically examine three possible models for the past evolution of
Wd~1: that it formed smaller, larger (and possibly bounced back to its current radius), or roughly the same size as it
is now. 
\emph{Our favoured model is that Wd~1 formed at a similar or lower density {\em to its current values}, with no primordial 
  mass segregation}.  This readily 
explains the low velocity dispersion of Wd~1, its lack of mass
segregation, and why the most massive stars are in regions of high
local surface density, and why Wd~1 is fairly smooth, but elongated \citep[see e.g.][]{2014MNRAS.438..620P}.  Any model in which Wd~1 was much denser in the past requires an element of fine-tuning in explaining why there is no signature of mass segregation (either
positive or negative), and if there is a signature why it matches the prediction from a static or collapsing model so well.

\section*{Acknowledgments}

RJP acknowledges support from the Royal Society in the form of a Dorothy Hodgkin Fellowship.

\bibliography{biblio_Wd1_MassSegr}

\begin{thebibliography}{}

\bibitem[\protect\citeauthoryear{{Adams}, {Hollenbach}, {Laughlin} \&
  {Gorti}}{{Adams} et~al.}{2004}]{2004ApJ...611..360A}
{Adams} F.~C.,  {Hollenbach} D.,  {Laughlin} G.,    {Gorti} U.,  2004, \apj,
  611, 360

\bibitem[\protect\citeauthoryear{{Allison} \& {Goodwin}}{{Allison} \&
  {Goodwin}}{2011}]{2011MNRAS.415.1967A}
{Allison} R.~J.,  {Goodwin} S.~P.,  2011, \mnras, 415, 1967

\bibitem[\protect\citeauthoryear{{Allison}, {Goodwin}, {Parker}, {de Grijs},
  {Portegies Zwart} \& {Kouwenhoven}}{{Allison}
  et~al.}{2009}]{2009ApJ...700L..99A}
{Allison} R.~J.,  {Goodwin} S.~P.,  {Parker} R.~J.,  {de Grijs} R.,  {Portegies
  Zwart} S.~F.,    {Kouwenhoven} M.~B.~N.,  2009, \apjl, 700, L99

\bibitem[\protect\citeauthoryear{{Allison}, {Goodwin}, {Parker}, {Portegies
  Zwart} \& {de Grijs}}{{Allison} et~al.}{2010}]{2010MNRAS.407.1098A}
{Allison} R.~J.,  {Goodwin} S.~P.,  {Parker} R.~J.,  {Portegies Zwart} S.~F.,
   {de Grijs} R.,  2010, \mnras, 407, 1098

\bibitem[\protect\citeauthoryear{{Allison}, {Goodwin}, {Parker}, {Portegies
  Zwart}, {de Grijs} \& {Kouwenhoven}}{{Allison} et~al.}{2009}]{allison09}
{Allison} R.~J.,  {Goodwin} S.~P.,  {Parker} R.~J.,  {Portegies Zwart} S.~F.,
  {de Grijs} R.,    {Kouwenhoven} M.~B.~N.,  2009, \mnras, 395, 1449

\bibitem[\protect\citeauthoryear{{Andr{\'e}}}{{Andr{\'e}}}{2002}]{2002Ap&SS.281...51A}
{Andr{\'e}} P.,  2002, \apss, 281, 51

\bibitem[\protect\citeauthoryear{{Ascenso}, {Alves} \& {Lago}}{{Ascenso}
  et~al.}{2009}]{2009Ap&SS.324..113A}
{Ascenso} J.,  {Alves} J.,    {Lago} M.~T.~V.~T.,  2009, \apss, 324, 113

\bibitem[\protect\citeauthoryear{{Banerjee} \& {Kroupa}}{{Banerjee} \&
  {Kroupa}}{2013}]{2013ApJ...764...29B}
{Banerjee} S.,  {Kroupa} P.,  2013, \apj, 764, 29

\bibitem[\protect\citeauthoryear{{Banerjee} \& {Kroupa}}{{Banerjee} \&
  {Kroupa}}{2015a}]{2015arXiv151004293B}
{Banerjee} S.,  {Kroupa} P.,  2015a, ArXiv e-prints

\bibitem[\protect\citeauthoryear{{Banerjee} \& {Kroupa}}{{Banerjee} \&
  {Kroupa}}{2015b}]{2015MNRAS.447..728B}
{Banerjee} S.,  {Kroupa} P.,  2015b, \mnras, 447, 728

\bibitem[\protect\citeauthoryear{{Bastian}, {Gieles}, {Ercolano} \&
  {Gutermuth}}{{Bastian} et~al.}{2009}]{2009MNRAS.392..868B}
{Bastian} N.,  {Gieles} M.,  {Ercolano} B.,    {Gutermuth} R.,  2009, \mnras,
  392, 868

\bibitem[\protect\citeauthoryear{{Baumgardt} \& {Kroupa}}{{Baumgardt} \&
  {Kroupa}}{2007}]{2007MNRAS.380.1589B}
{Baumgardt} H.,  {Kroupa} P.,  2007, \mnras, 380, 1589

\bibitem[\protect\citeauthoryear{{Bonanos}}{{Bonanos}}{2007}]{2007AJ....133.2696B}
{Bonanos} A.~Z.,  2007, \aj, 133, 2696

\bibitem[\protect\citeauthoryear{{Bonnell}, {Bate}, {Clarke} \&
  {Pringle}}{{Bonnell} et~al.}{1997}]{1997MNRAS.285..201B}
{Bonnell} I.~A.,  {Bate} M.~R.,  {Clarke} C.~J.,    {Pringle} J.~E.,  1997,
  \mnras, 285, 201

\bibitem[\protect\citeauthoryear{{Bonnell}, {Bate}, {Clarke} \&
  {Pringle}}{{Bonnell} et~al.}{2001}]{2001MNRAS.323..785B}
{Bonnell} I.~A.,  {Bate} M.~R.,  {Clarke} C.~J.,    {Pringle} J.~E.,  2001,
  \mnras, 323, 785

\bibitem[\protect\citeauthoryear{{Brandner}, {Clark}, {Stolte}, {Waters},
  {Negueruela} \& {Goodwin}}{{Brandner} et~al.}{2008}]{Brandner:2008lr}
{Brandner} W.,  {Clark} J.~S.,  {Stolte} A.,  {Waters} R.,  {Negueruela} I.,
  {Goodwin} S.~P.,  2008, \aap, 478, 137

\bibitem[\protect\citeauthoryear{{Bressert}, {Bastian}, {Gutermuth}, {Megeath},
  {Allen}, {Evans} II, {Rebull}, {Hatchell}, {Johnstone}, {Bourke}, {Cieza},
  {Harvey}, {Merin}, {Ray} \& {Tothill}}{{Bressert}
  et~al.}{2010}]{2010MNRAS.409L..54B}
{Bressert} E.,  {Bastian} N.,  {Gutermuth} R.,  {Megeath} S.~T.,  {Allen} L.,
  {Evans} II N.~J.,  {Rebull} L.~M.,  {Hatchell} J.,  {Johnstone} D.,  {Bourke}
  T.~L.,  {Cieza} L.~A.,  {Harvey} P.~M.,  {Merin} B.,  {Ray} T.~P.,
  {Tothill} N.~F.~H.,  2010, \mnras, 409, L54

\bibitem[\protect\citeauthoryear{{Caputo}, {de Vries} \& {Portegies
  Zwart}}{{Caputo} et~al.}{2014}]{2014MNRAS.445..674C}
{Caputo} D.~P.,  {de Vries} N.,    {Portegies Zwart} S.,  2014, \mnras, 445,
  674

\bibitem[\protect\citeauthoryear{{Cardelli}, {Clayton} \& {Mathis}}{{Cardelli}
  et~al.}{1989}]{1989ApJ...345..245C}
{Cardelli} J.~A.,  {Clayton} G.~C.,    {Mathis} J.~S.,  1989, \apj, 345, 245

\bibitem[\protect\citeauthoryear{{Cartwright} \& {Whitworth}}{{Cartwright} \&
  {Whitworth}}{2004}]{2004MNRAS.348..589C}
{Cartwright} A.,  {Whitworth} A.~P.,  2004, \mnras, 348, 589

\bibitem[\protect\citeauthoryear{{Clark}, {Negueruela}, {Crowther} \&
  {Goodwin}}{{Clark} et~al.}{2005}]{2005A&A...434..949C}
{Clark} J.~S.,  {Negueruela} I.,  {Crowther} P.~A.,    {Goodwin} S.~P.,  2005,
  \aap, 434, 949

\bibitem[\protect\citeauthoryear{{Cottaar}, {Meyer}, {Andersen} \&
  {Espinoza}}{{Cottaar} et~al.}{2012}]{cottaar12}
{Cottaar} M.,  {Meyer} M.~R.,  {Andersen} M.,    {Espinoza} P.,  2012, \aap,
  539, A5

\bibitem[\protect\citeauthoryear{{Degl'Innocenti}, {Prada Moroni}, {Marconi} \&
  {Ruoppo}}{{Degl'Innocenti} et~al.}{2008}]{DeglInnocenti:2008fk}
{Degl'Innocenti} S.,  {Prada Moroni} P.~G.,  {Marconi} M.,    {Ruoppo} A.,
  2008, \apss, 316, 25

\bibitem[\protect\citeauthoryear{{Devillard}}{{Devillard}}{2001}]{Devillard:2001fk}
{Devillard} N.,  2001, in {F.~R.~Harnden Jr., F.~A.~Primini, \& H.~E.~Payne}
  ed., Astronomical Data Analysis Software and Systems X, Vol.~238 of
  Astronomical Society of the Pacific Conference Series, {ESO C Library for an
  Image Processing Software Environment (eclipse)}.
p.~525

\bibitem[\protect\citeauthoryear{{Diolaiti}, {Bendinelli}, {Bonaccini},
  {Close}, {Currie} \& {Parmeggiani}}{{Diolaiti}
  et~al.}{2000}]{2000A&AS..147..335D}
{Diolaiti} E.,  {Bendinelli} O.,  {Bonaccini} D.,  {Close} L.,  {Currie} D.,
  {Parmeggiani} G.,  2000, \aaps, 147, 335

\bibitem[\protect\citeauthoryear{{Gennaro}, {Brandner}, {Stolte} \&
  {Henning}}{{Gennaro} et~al.}{2011}]{gennaro11}
{Gennaro} M.,  {Brandner} W.,  {Stolte} A.,    {Henning} T.,  2011, \mnras,
  412, 2469

\bibitem[\protect\citeauthoryear{{Gennaro}, {Prada Moroni} \&
  {Tognelli}}{{Gennaro} et~al.}{2012}]{2012MNRAS.420..986G}
{Gennaro} M.,  {Prada Moroni} P.~G.,    {Tognelli} E.,  2012, \mnras, 420, 986

\bibitem[\protect\citeauthoryear{{Gieles}, {Moeckel} \& {Clarke}}{{Gieles}
  et~al.}{2012}]{2012MNRAS.426L..11G}
{Gieles} M.,  {Moeckel} N.,    {Clarke} C.~J.,  2012, \mnras, 426, L11

\bibitem[\protect\citeauthoryear{{Goodwin}}{{Goodwin}}{2009}]{2009Ap&SS.324..259G}
{Goodwin} S.~P.,  2009, \apss, 324, 259

\bibitem[\protect\citeauthoryear{{Goodwin} \& {Bastian}}{{Goodwin} \&
  {Bastian}}{2006}]{2006MNRAS.373..752G}
{Goodwin} S.~P.,  {Bastian} N.,  2006, \mnras, 373, 752

\bibitem[\protect\citeauthoryear{{Kroupa}}{{Kroupa}}{1995}]{1995MNRAS.277.1491K}
{Kroupa} P.,  1995, \mnras, 277, 1491

\bibitem[\protect\citeauthoryear{{Kruijssen}, {Pelupessy}, {Lamers}, {Portegies
  Zwart} \& {Icke}}{{Kruijssen} et~al.}{2011}]{2011MNRAS.414.1339K}
{Kruijssen} J.~M.~D.,  {Pelupessy} F.~I.,  {Lamers} H.~J.~G.~L.~M.,  {Portegies
  Zwart} S.~F.,    {Icke} V.,  2011, \mnras, 414, 1339

\bibitem[\protect\citeauthoryear{{Krumholz}, {Klein} \& {McKee}}{{Krumholz}
  et~al.}{2007}]{2007ApJ...656..959K}
{Krumholz} M.~R.,  {Klein} R.~I.,    {McKee} C.~F.,  2007, \apj, 656, 959

\bibitem[\protect\citeauthoryear{{K{\"u}pper}, {Maschberger}, {Kroupa} \&
  {Baumgardt}}{{K{\"u}pper} et~al.}{2011}]{2011MNRAS.417.2300K}
{K{\"u}pper} A.~H.~W.,  {Maschberger} T.,  {Kroupa} P.,    {Baumgardt} H.,
  2011, \mnras, 417, 2300

\bibitem[\protect\citeauthoryear{{Lada}}{{Lada}}{2010}]{2010RSPTA.368..713L}
{Lada} C.~J.,  2010, Royal Society of London Philosophical Transactions Series
  A, 368, 713

\bibitem[\protect\citeauthoryear{{Lada} \& {Lada}}{{Lada} \&
  {Lada}}{2003}]{2003ARA&A..41...57L}
{Lada} C.~J.,  {Lada} E.~A.,  2003, \araa, 41, 57

\bibitem[\protect\citeauthoryear{{Lada}, {Margulis} \& {Dearborn}}{{Lada}
  et~al.}{1984}]{1984ApJ...285..141L}
{Lada} C.~J.,  {Margulis} M.,    {Dearborn} D.,  1984, \apj, 285, 141

\bibitem[\protect\citeauthoryear{{Longmore}, {Kruijssen}, {Bastian}, {Bally},
  {Rathborne}, {Testi}, {Stolte}, {Dale}, {Bressert} \& {Alves}}{{Longmore}
  et~al.}{2014}]{2014prpl.conf..291L}
{Longmore} S.~N.,  {Kruijssen} J.~M.~D.,  {Bastian} N.,  {Bally} J.,
  {Rathborne} J.,  {Testi} L.,  {Stolte} A.,  {Dale} J.,  {Bressert} E.,
  {Alves} J.,  2014, Protostars and Planets VI, pp 291--314

\bibitem[\protect\citeauthoryear{{Lopez}, {Krumholz}, {Bolatto}, {Prochaska} \&
  {Ramirez-Ruiz}}{{Lopez} et~al.}{2011}]{2011ApJ...731...91L}
{Lopez} L.~A.,  {Krumholz} M.~R.,  {Bolatto} A.~D.,  {Prochaska} J.~X.,
  {Ramirez-Ruiz} E.,  2011, \apj, 731, 91

\bibitem[\protect\citeauthoryear{{Marigo}, {Girardi}, {Bressan}, {Groenewegen},
  {Silva} \& {Granato}}{{Marigo} et~al.}{2008}]{Marigo:2008lr}
{Marigo} P.,  {Girardi} L.,  {Bressan} A.,  {Groenewegen} M.~A.~T.,  {Silva}
  L.,    {Granato} G.~L.,  2008, \aap, 482, 883

\bibitem[\protect\citeauthoryear{{Maschberger} \& {Clarke}}{{Maschberger} \&
  {Clarke}}{2011}]{2011MNRAS.416..541M}
{Maschberger} T.,  {Clarke} C.~J.,  2011, \mnras, 416, 541

\bibitem[\protect\citeauthoryear{{Moeckel}, {Holland}, {Clarke} \&
  {Bonnell}}{{Moeckel} et~al.}{2012}]{2012MNRAS.425..450M}
{Moeckel} N.,  {Holland} C.,  {Clarke} C.~J.,    {Bonnell} I.~A.,  2012,
  \mnras, 425, 450

\bibitem[\protect\citeauthoryear{{Negueruela}, {Clark} \&
  {Ritchie}}{{Negueruela} et~al.}{2010}]{Negueruela:2010fj}
{Negueruela} I.,  {Clark} J.~S.,    {Ritchie} B.~W.,  2010, \aap, 516, A78

\bibitem[\protect\citeauthoryear{{Nishiyama}, {Nagata}, {Kusakabe},
  {Matsunaga}, {Naoi}, {Kato}, {Nagashima}, {Sugitani}, {Tamura}, {Tanab{\'e}}
  \& {Sato}}{{Nishiyama} et~al.}{2006}]{2006ApJ...638..839N}
{Nishiyama} S.,  {Nagata} T.,  {Kusakabe} N.,  {Matsunaga} N.,  {Naoi} T.,
  {Kato} D.,  {Nagashima} C.,  {Sugitani} K.,  {Tamura} M.,  {Tanab{\'e}} T.,
   {Sato} S.,  2006, \apj, 638, 839

\bibitem[\protect\citeauthoryear{{Parker}}{{Parker}}{2014}]{2014MNRAS.445.4037P}
{Parker} R.~J.,  2014, \mnras, 445, 4037

\bibitem[\protect\citeauthoryear{{Parker}, {Bouvier}, {Goodwin}, {Moraux},
  {Allison}, {Guieu} \& {G{\"u}del}}{{Parker}
  et~al.}{2011}]{2011MNRAS.412.2489P}
{Parker} R.~J.,  {Bouvier} J.,  {Goodwin} S.~P.,  {Moraux} E.,  {Allison}
  R.~J.,  {Guieu} S.,    {G{\"u}del} M.,  2011, \mnras, 412, 2489

\bibitem[\protect\citeauthoryear{{Parker} \& {Dale}}{{Parker} \&
  {Dale}}{2013}]{2013MNRAS.432..986P}
{Parker} R.~J.,  {Dale} J.~E.,  2013, \mnras, 432, 986

\bibitem[\protect\citeauthoryear{{Parker} \& {Dale}}{{Parker} \&
  {Dale}}{2017}]{2017MNRAS.470..390P}
{Parker} R.~J.,  {Dale} J.~E.,  2017, \mnras, 470, 390

\bibitem[\protect\citeauthoryear{{Parker}, {Dale} \& {Ercolano}}{{Parker}
  et~al.}{2015}]{2015MNRAS.446.4278P}
{Parker} R.~J.,  {Dale} J.~E.,    {Ercolano} B.,  2015, \mnras, 446, 4278

\bibitem[\protect\citeauthoryear{{Parker} \& {Goodwin}}{{Parker} \&
  {Goodwin}}{2015}]{2015MNRAS.449.3381P}
{Parker} R.~J.,  {Goodwin} S.~P.,  2015, \mnras, 449, 3381

\bibitem[\protect\citeauthoryear{{Parker}, {Maschberger} \& {Alves de
  Oliveira}}{{Parker} et~al.}{2012}]{2012MNRAS.426.3079P}
{Parker} R.~J.,  {Maschberger} T.,    {Alves de Oliveira} C.,  2012, \mnras,
  426, 3079

\bibitem[\protect\citeauthoryear{{Parker} \& {Meyer}}{{Parker} \&
  {Meyer}}{2012}]{2012MNRAS.427..637P}
{Parker} R.~J.,  {Meyer} M.~R.,  2012, \mnras, 427, 637

\bibitem[\protect\citeauthoryear{{Parker} \& {Quanz}}{{Parker} \&
  {Quanz}}{2012}]{2012MNRAS.419.2448P}
{Parker} R.~J.,  {Quanz} S.~P.,  2012, \mnras, 419, 2448

\bibitem[\protect\citeauthoryear{{Parker} \& {Wright}}{{Parker} \&
  {Wright}}{2016}]{2016MNRAS.457.3430P}
{Parker} R.~J.,  {Wright} N.~J.,  2016, \mnras, 457, 3430

\bibitem[\protect\citeauthoryear{{Parker}, {Wright}, {Goodwin} \&
  {Meyer}}{{Parker} et~al.}{2014}]{2014MNRAS.438..620P}
{Parker} R.~J.,  {Wright} N.~J.,  {Goodwin} S.~P.,    {Meyer} M.~R.,  2014,
  \mnras, 438, 620

\bibitem[\protect\citeauthoryear{{Peretto}, {Andr{\'e}} \&
  {Belloche}}{{Peretto} et~al.}{2006}]{2006A&A...445..979P}
{Peretto} N.,  {Andr{\'e}} P.,    {Belloche} A.,  2006, \aap, 445, 979

\bibitem[\protect\citeauthoryear{{Pfalzner} \& {Kaczmarek}}{{Pfalzner} \&
  {Kaczmarek}}{2013}]{2013A&A...559A..38P}
{Pfalzner} S.,  {Kaczmarek} T.,  2013, \aap, 559, A38

\bibitem[\protect\citeauthoryear{{Skrutskie}, {Cutri}, {Stiening} \& {et
  al.}}{{Skrutskie} et~al.}{2006}]{Skrutskie:2006uq}
{Skrutskie} M.~F.,  {Cutri} R.~M.,  {Stiening} R.,    {et al.} 2006, \aj, 131,
  1163

\bibitem[\protect\citeauthoryear{{Smith}, {Goodwin}, {Fellhauer} \&
  {Assmann}}{{Smith} et~al.}{2013}]{2013MNRAS.428.1303S}
{Smith} R.,  {Goodwin} S.,  {Fellhauer} M.,    {Assmann} P.,  2013, \mnras,
  428, 1303

\bibitem[\protect\citeauthoryear{{Stetson}}{{Stetson}}{1987}]{Stetson:1987qy}
{Stetson} P.~B.,  1987, \pasp, 99, 191

\bibitem[\protect\citeauthoryear{{Trani}, {Mapelli} \& {Bressan}}{{Trani}
  et~al.}{2014}]{2014MNRAS.445.1967T}
{Trani} A.~A.,  {Mapelli} M.,    {Bressan} A.,  2014, \mnras, 445, 1967

\bibitem[\protect\citeauthoryear{{Tutukov}}{{Tutukov}}{1978}]{1978A&A....70...57T}
{Tutukov} A.~V.,  1978, \aap, 70, 57

\bibitem[\protect\citeauthoryear{{Vincke}, {Breslau} \& {Pfalzner}}{{Vincke}
  et~al.}{2015}]{2015A&A...577A.115V}
{Vincke} K.,  {Breslau} A.,    {Pfalzner} S.,  2015, \aap, 577, A115

\bibitem[\protect\citeauthoryear{{Westerlund}}{{Westerlund}}{1961}]{1961AJ.....66T..57W}
{Westerlund} B.,  1961, \aj, 66, 57

\bibitem[\protect\citeauthoryear{{Wright}, {Parker}, {Goodwin} \&
  {Drake}}{{Wright} et~al.}{2014}]{2014MNRAS.438..639W}
{Wright} N.~J.,  {Parker} R.~J.,  {Goodwin} S.~P.,    {Drake} J.~J.,  2014,
  \mnras, 438, 639

\bibitem[\protect\citeauthoryear{{Zinnecker} \& {Yorke}}{{Zinnecker} \&
  {Yorke}}{2007}]{2007ARA&A..45..481Z}
{Zinnecker} H.,  {Yorke} H.~W.,  2007, \araa, 45, 481

\end{thebibliography}
\bibliographystyle{mn2e}

\label{lastpage}

\end{document}